\documentclass[journal,twoside,web]{ieeecolor}
\usepackage{tmi}
\usepackage{cite}
\usepackage{amsmath,amssymb,amsfonts}
\usepackage{algorithmic}
\usepackage{graphicx}
\usepackage{textcomp}
\usepackage{amsfonts}
\usepackage{amssymb}
\usepackage{multirow}
\usepackage{bm}

\def\BibTeX{{\rm B\kern-.05em{\sc i\kern-.025em b}\kern-.08em
    T\kern-.1667em\lower.7ex\hbox{E}\kern-.125emX}}
\begin{document}
\title{Confidence-guided Lesion Mask-based Simultaneous Synthesis of Anatomic and Molecular MR Images in Patients with Post-treatment Malignant Gliomas}
%
%
%

\author{Pengfei Guo,~\IEEEmembership{Student Member,~IEEE,}
	Puyang Wang,~\IEEEmembership{Student Member,~IEEE,}
	Rajeev Yasarla,~\IEEEmembership{Student Member,~IEEE,} Jinyuan Zhou,
	Vishal M. Patel,~\IEEEmembership{Senior Member,~IEEE,}
	and Shanshan Jiang,~\IEEEmembership{Member,~IEEE} 
	\thanks{Pengfei Guo, Puyang Wang, Rajeev Yasarla, and Vishal M. Patel are with the Whiting School of Engineering, Johns Hopkins University,
		(e-mail: \{pguo4, pwang47, ryasarl1, vpatel36\}@jhu.edu).}
	\thanks{Jinyuan Zhou and Shanshan Jiang are with the School of Medicine, Johns Hopkins University, (e-mail:
		\{jzhou2,sjiang21\}@jhmi.edu).}
	}

\maketitle


\begin{abstract}
	Data-driven automatic approaches have demonstrated their great potential in resolving various clinical
	diagnostic dilemmas in neuro-oncology, especially with
	the help of standard anatomic and advanced molecular MR images.
	However, data quantity and quality remain a key determinant of, and a significant limit on, the potential of such applications. In our previous work, we explored synthesis of
	anatomic and molecular MR image network (SAMR) in patients with post-treatment malignant glioms.
	Now, we extend it and propose Confidence Guided SAMR (CG-SAMR) that synthesizes data from lesion information to multi-modal anatomic sequences, including T1-weighted ($T_1$w), gadolinium  enhanced $T_1$w (Gd-$T_1$w), T2-weighted ($T_2$w), and fluid-attenuated inversion recovery ($FLAIR$), and the molecular amide proton transfer‐weighted ($APT$w) sequence. We introduce a module which guides the synthesis based on confidence measure about the intermediate results. Furthermore, we extend the proposed architecture for unsupervised synthesis so that unpaired data can be used for training the network. Extensive experiments on real clinical data demonstrate that the proposed model can perform better than the state-of-the-art synthesis methods.
\end{abstract}
\begin{figure}[]
	\centering
	\includegraphics[width=.9\columnwidth]{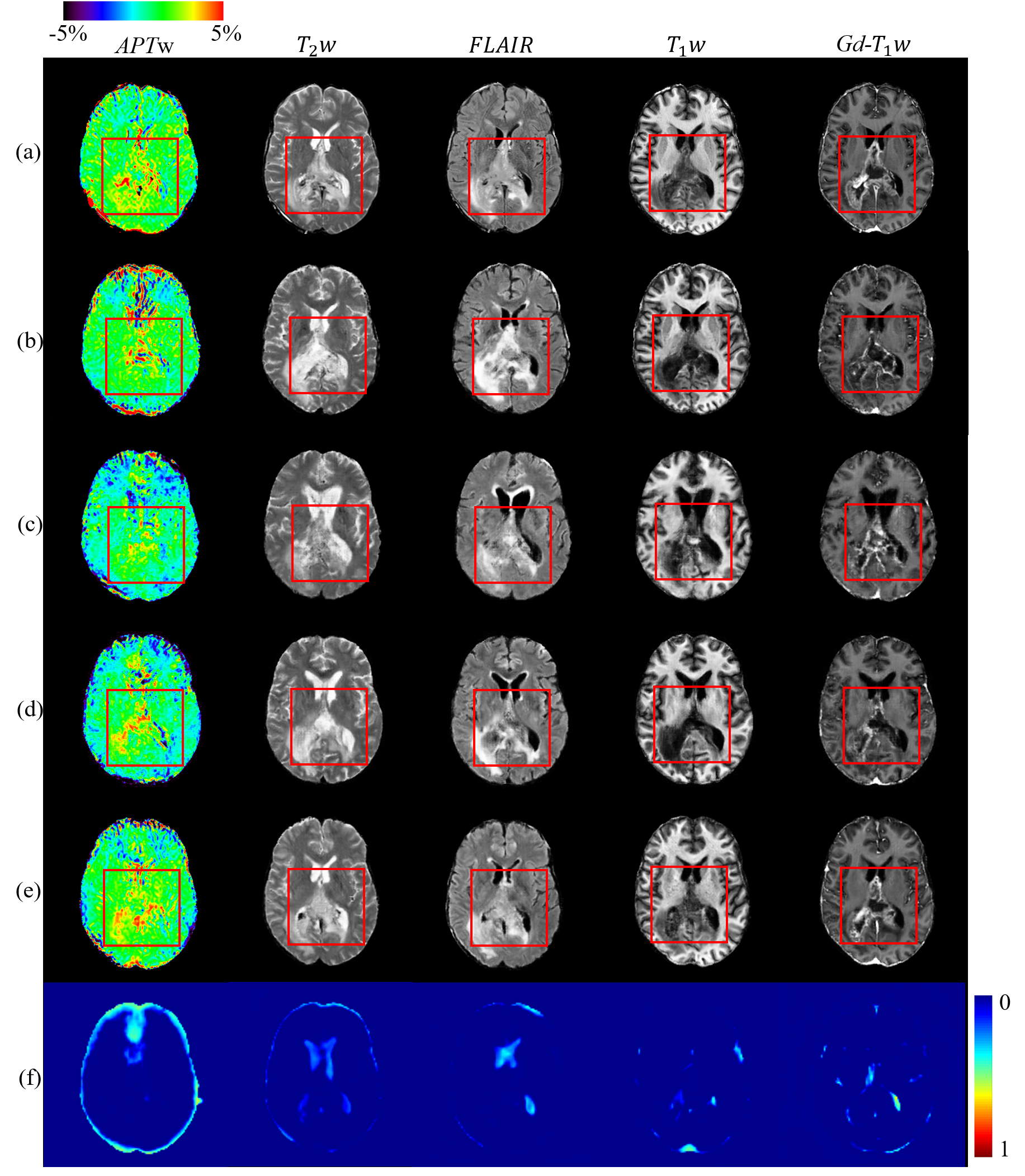}
	\vskip-0.2cm
	\caption{Sample multi-modal MR image synthesis results. (a) Ground truth. (b) Synthesis using Shin et al.~\cite{au14}.  (c) Synthesis using pix2pix~\cite{au8}. (d) Synthesis using pix2pixHD~\cite{au9}. (e) Synthesis using the proposed CG-SAMR method. (f) Confidence maps from CG-SAMR, where blue represents the most confident and red represents the least confident region. \label{fig1}}
	\vskip-0.5cm
\end{figure}

\begin{IEEEkeywords}
	Molecular MRI, Multi-modal Synthesis, GAN.
\end{IEEEkeywords}

%
\IEEEpeerreviewmaketitle

\section{Introduction}

Glioblastoma (GBM) is the most malignant and frequently occurring type of primary brain tumor in adult. Despite the development of various aggressive treatments, patients with GBM inevitably suffers tumor recurrence with an extremely poor prognosis \cite{au2}. The dilemma in the clinical management of post-treatment patients remains precise assessment of  the treatment responsiveness. However, it is mostly relied on  pathological evaluations of biopsies~\cite{au1}. Magnetic resonance imaging (MRI) is considered the best non-invasive assessment method of GBM treatment responsiveness~\cite{au62}. Compared with anatomic MRI, such as T1-weighted ($T_1$w), gadolinium enhanced $T_1$w (Gd-$T_1$w), T2-weighted ($T_2$w), and fluid-attenuated inversion recovery ($FLAIR$) images, amide proton transfer-weighted ($APT$w) MRI is a novel molecular imaging technique. It has been proved to positively influence the clinical management by different labs across the world~\cite{au60}. Recently, convolutional neural network (CNN) based medical image analysis methods have provided exciting solutions in neuro-oncologic community~\cite{au3}. Several studies have demonstrated that CNN-based methods outperform humans on fine-grained classification tasks but require a large amount of accurately annotated data with rich diversity~\cite{au61}. Compared with demanding large anatomic MRI datasets, it becomes even more impractical when collecting cutting edge MR image data. Furthermore, obtaining aligned lesion annotations on the corresponding co-registrated multi-modal MR images (namely, paired training data) is costly, since expert radiologists are required to label and verify the data. While deploying conventional data augmentations, such as rotation, flipping, random cropping, and distortion, during training partly mitigates such issues, the performance of CNN models is still limited by the diversity of the dataset~\cite{au4}. In this paper, we address the problem of synthesizing meaningful high quality anatomic $T_1$w, Gd-$T_1$w, $T_2$w, $FLAIR$, and molecular $APT$w MR images based on the input lesion information.

\indent Goodfellow et al. \cite{au7} proposed the generative adversarial networks (GAN) and first applied to synthesize photo-realistic images. Isola et al. \cite{au8} and Wang et al. \cite{au9} further investigated conditional GAN and achieved impressive solution to image-to-image translation problems. Synthesizing realistic MR images is a difficult task since radiographic features dramatically varies on MR images corresponding to underlying diverse pathological changes. Nevertheless, several generative models have been successfully proposed for MRI synthesis. Nguyen et al. \cite{au10} and Chartsias et al. \cite{au12} proposed CNN-based architectures to synthesize cross-modality MR images. Cordier et al. \cite{au13} further used a generative model for multi-modal MR images with brain tumors from a single label map. However, their inputs are conventional MRI modalities, and the diversity of the synthesized images is limited by the training images. Moreover, the method is not yet capable of producing manipulated outputs. Shin et al. \cite{au14} adopted Pix2Pix \cite{au8} to transfer brain anatomy and lesion segmentation maps to multi-modal MR images with brain tumors. Although, their approach can synthesize realistic brain anatomy for multiple MRI sequences, it does not consider significant differences of radiographic features between anatomic and molecular MRI. Moreover, pathological information are high frequency components and may need extra supervision during synthesis. As a result, their method cannot produce realistic molecular MR images and fails around the lesion region (see Figure~\ref{fig1}(b)).

In our previous work, synthesis of anatomic and molecular MR images network (SAMR)~\cite{au59}, a novel generative model was proposed to simultaneously synthesize a diverse set of anatomic and molecular MR images. It takes arbitrarily manipulated lesion masks as input, which is facilitated by brain atlas generated from training data. SAMR \cite{au59} is a GAN-based approach, which consists of a stretch-out up-sampling module, a segmentation consistency module, and multi-scale label-wise discriminators. In this paper, we extend SAMR \cite{au59} by incorporating extra supervision on the latent features and their confidence information to further improve the synthetic performance. Intuitively, directly providing the estimated synthesized images (i.e. intermediate results) to the subsequent layers of the network may propagate errors to the final synthesized images. With the confidence map module, the proposed algorithm is capable to measure an uncertainty metric of the intermediate results and block the flow of incorrect estimation. To this end, we formulate a joint task of estimating the confidence score at each pixel location of intermediate results and synthesizing realistic multi-modal MR images. Figure~\ref{fig1}(e) presents sample results from proposed network, where CG-SAMR generates realistic multi-modal brain MR images with more detailed pathological information as compared with Figure~\ref{fig1}(b-d). Furthermore, to overcome the insufficiency of paired training data, we modify the network to allow unsupervised training, namely unpaired CG-SAMR (UCG-SAMR). In other words, the proposed unsupervised approach does not require aligned pairs of lesion segmentation maps and multi-modal MR images during training. This is achieved by adding an extra GAN which reverses the synthesis process to a segmentation task. In summary, this paper makes the following contributions:
\begin{itemize}
	\item A novel GAN-based model, called CG-SAMR, is proposed to synthesize high quality multi-modal anatomic and molecular MR images with controllable lesion information.
	\item A novel stretch-out up-sampling module in the decoder is proposed which performs customized synthesis for images of each MR sequence.
	\item Confidence scores of each sequence measured during synthesis are used to guide the subsequent layers for better synthetic performance.
	\item Multi-scale label-wise discriminators are developed to provide specific supervision on distinguishing region of interests (ROIs).
	\item In order to increase the diversity of the synthesized data, rather than explicitly using white matter (WM), gray matter (GM), and cerebrospinal fluid (CSF) masks, we leverage atlas of each sequence to provide brain anatomic information in CG-SAMR.
	\item We demonstrate the feasibility of extending the application of CG-SAMR network to unpaired data training.
	\item Comparisons have been performed against several recent state-of-the-art paired/unpaired synthesis approaches. Furthermore, an ablation study is conducted to demonstrate the improvements obtained by various components of the proposed method.
\end{itemize}
Rest of the paper is organized as follows. Section~\ref{sec2} provides a review of some related works.   Details of the proposed method
are given in Section~\ref{sec3}. Implementation details, experimental results, and ablation study are given in Section~\ref{sec4}. Finally, Section~\ref{sec5} concludes the paper with a brief
discussion and summary.

\begin{figure*}
	\centering
	\includegraphics[width=.8\textwidth]{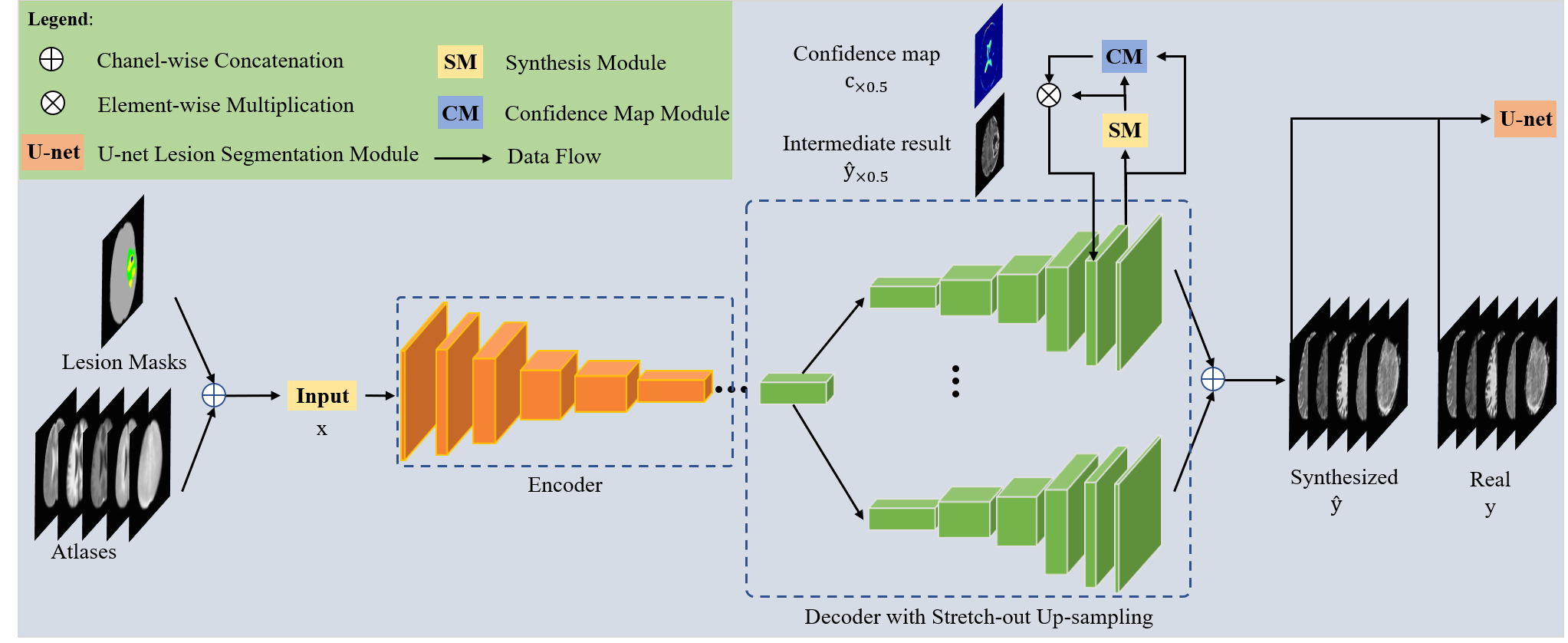}
	\caption{An overview of the proposed CG-SAMR network. The goal of the CG-SAMR network is to produce realistic multi-model MR images given
		the corresponding lesion masks and atlases. The orange blocks indicate the encoder part. The green blocks represent the decoder part with stretch-out up-sampling, in which we leverage same latent feature maps to perform customized synthesis for each MRI sequence. The synthesis module produces the intermediate results for each branch of stretch-out up-sampling and is denoted as SM. CM represents the confidence map module that computes confidence maps to guide
		the subsequent networks. The U-net lesion segmentation module regularize encoder-decoder part to produce lesion regions with correct radiographic features by a lesion shape consistency loss $\mathcal{L}_{\text{SC}}$. \label{fig2}}
	\vskip-0.5cm
\end{figure*}

\section{Related Works}
\label{sec2}
The goal of MR images synthesis is to generate target images with realistic radiographic features~\cite{au23}. MR image synthesis technically can be achieved by a generative model that translates the source domain to the MR image domain. The source domain usually belongs to noise or different modalities/contrast types (e.g., from CT images to MR images and from $T_1$w images to $T_2$w images). In what follows, we review some recent studies on this topic and applications of modeling uncertainty in CNN.\\

\subsection{Conventional Methods}
\indent The conventional medical image synthesis methods include intensity-based methods and registration-based methods~\cite{au24}. Intensity-based methods essentially learn a transformation function mapping source intensities to target intensities. Roy \emph{et al.}~\cite{au25} proposed an example-based approach relying on sparse reconstruction from image patches to achieve contrast synthesis and further extended it under the setting of patch-based compressed sensing~\cite{au27}. Joy \emph{et al.}~\cite{au26} leveraged random forest regression to learn the nonlinear intensity mappings for synthesizing full-head $T_2$w images and $FLAIR$ images. Huang \emph{et al.}~\cite{au28} proposed a geometry regularized joint dictionary learning framework to synthesize cross-modality MR images. For registration-based methods, the synthesized images are generated by the registration between a source images and target co-registered images~\cite{au29}. Cardoso \emph{et al.}~\cite{au30} further extended this idea to synthesize expected intensities in an unseen image modality by a template-based multi-modal generative mixture-model.
\subsection{CNN-based Methods}
 With the development of deep learning, CNN-based medical image synthesis methods have shown significant improvements over the conventional methods. Instead of using patch-based methods~\cite{au35,au36}, Sevetlidis \emph{et al.}~\cite{au34} introduced a whole image synthesis approach relying on a CNN-based autoencoder architecture. Nguyen \emph{et al.}~\cite{au10} and Chartsias \emph{et al.}~\cite{au12} proposed CNN-based architectures integrating intensity features from images to synthesize cross-modality MR images. Various GAN-based methods have also been used for medical image analysis~\cite{au37,au38}. Shin et al.~\cite{au14} adopted pix2pix~\cite{au8} to transfer brain anatomy and lesion segmentation maps to multi-modal MR images with brain tumors. It shows the benefit of using brain anatomy prior, such as WM, GM, CSF masks, to facilitate MR image synthesis. 
 
  One major challenge of image synthesis is that paired source/target images are required during training which are expensive to acquire. Recent developments in GAN-based architectures, cycle-consistent adversarial networks (CycleGAN)~\cite{au39} and unsupervised image-to-image translation networks (UNIT)~\cite{au48} point to a promising direction for cross-modality biomedical image synthesis using unpaired source/target images). Wolterink \emph{et al.}~\cite{au40} leveraged cycle consistency to achieve bidirectional MR/CT images synthesis. Chartsias \emph{et al.}~\cite{au41} proposed a two stage framework for MR/CT images synthesis and demonstrated that the synthesized data can further improve the segmentation performance. Zhang \emph{et al.}~\cite{au42} and Huo \emph{et al.}~\cite{au43} introduced SynSeg-Net to achieve bidirectional synthesis and anatomy segmentation. In their approach, the source domain is the MR images as well as segmentation labels and the target domain is CT images. Inspired by these works, we also add an extra GAN-based network to CG-SAMR and leverage cycle consistency to allow the training using unpaired data.
 
 \subsection{Modeling Uncertainty in CNN}
 Many recent approaches model the uncertainty and use it to benefit the network on different applications.
 Kendall \emph{et al.}~\cite{au44} leveraged the Bayesian deep learning models to demonstrate the benefit of modeling uncertainty on  semantic segmentation and depth regression tasks. In~\cite{au45}, Kendall \emph{et al.} extended the previous work~\cite{au44} to multi-task learning by proposing a multi-task loss function maximizing the Gaussian likelihood with homoscedastic
 uncertainty. Yasarla \emph{et al.}~\cite{au46} and Jose \emph{et al.}~\cite{au51} modeled the aleatoric uncertainty as maximum likelihood inference on image restoration and ultrasound image segmentation tasks, respectively. Inspired by these works, we introduce a novel loss function to measure the confidence score of the intermediate synthesis results and guide the subsequent networks of CG-SAMR by the estimated confidence scores.

\section{Methodology}
\label{sec3}
Figure~\ref{fig2} gives an overview of the proposed encoder and decoder part in CG-SAMR framework. By incorporating multi-scale label-wise discriminators and shape consistency-based optimization, the generator aims to produce meaningful high-quality anatomical and molecular MR images with diverse and controllable lesion information. While applying 3D convolution operations might reflect the reality of data, the output of the proposed method is multi-modal MRI image slices, since voxel size between anatomical and molecular MRI in axial direction is significantly different and re-sampling to isotropic resolution can severely degrade the image quality. Detailed imaging parameters are given in Section~\ref{sec:da}. In what follows, we describe key parts of the network and training processes using paired and unpaired data.\\
\begin{figure}
	\centering
	\includegraphics[width=.8\columnwidth]{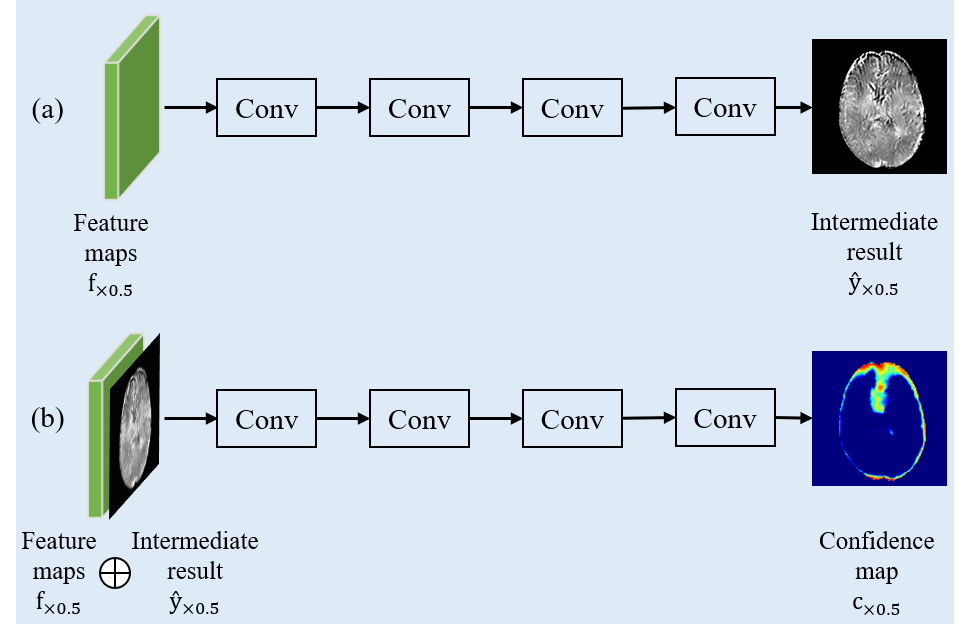}
	\caption{(a) Synthesis module. (b) Confidence map module. Here, Conv represents a convolution block that contains a convolutional layer, a batch normalization layer, and a Rectified Linear Units (ReLU) activation. $\oplus$ is the channel-wise concatenation. \label{fig3}}
\end{figure}
\subsection{Multi-modal MRI Generation}
Our generator architecture is inspired by the models proposed by Johnson et al. \cite{au15} and Wang et al. \cite{au9}. The generator network, consists of two components (see Fig.~\ref{fig1}): an encoder and a decoder with stretch-out up-sampling module. Let the set of multi-model MR images be denoted as $\mathcal{Y}$ and the corresponding set of lesion segmentation maps and anatomic prior as $\mathcal{X}$. The generator aims to synthesize multi-modal MR images $y \in \mathcal{Y}$ given input $x \in \mathcal{X}$. Unlike many deep learning-based methods that directly
synthesize MR images from input, we first estimate the
intermediate synthesis results $\hat{y}_{\times 0.5}$ (0.5 scale size of $y$) and the corresponding confidence map $c_{\times 0.5}$, then use them to guide the synthesis of the final output $\hat{y}$. The input $x$ is passed through the encoder module to get the latent feature maps. Then, the same latent feature maps are passed through each branch of the stretch-out up-sampling block to perform customized synthesis.

The encoder part (orange blocks in Figure~\ref{fig2}) consists of a fully-convolutional module with 5 layers and subsequent 3 residual learning blocks (ResBlock) \cite{au49}. We set the kernel size and stride equal to 7 and 1, respectively, for the first layer. For the purpose of down-sampling, instead of using maximum-pooling, the stride of other 4 layers is set equal to 2. Rectified Linear Unit (ReLu) activation and batch normalization are sequentially added after each layer. To learn better transformation functions and representations through a deeper perception, the depth of the encoder network is increased by 3 ResBlocks \cite{au4, au49}. We can observe the significant different radiographic features between anatomic and molecular MR images as shown in Figure~\ref{fig1}(a), which vastly increases the difficulty of simultaneous synthesis. To address this issue, the decoder part (green blocks in Figure~\ref{fig2}) consists of 3 ResBlocks and a stretch-out up-sampling module that contains 5 same sub-modules designed to utilize the same latent representations from the preceding ResBlock and perform customized synthesis for each MR sequence. Each sub-module contains a symmetric architecture with a fully-convolutional module in the encoder. All convolutional layers are replaced by transposed convolutional layers for up-sampling. The synthesized multi-modal MR images are produced from each sub-model. 
\subsection{Synthesis and Confidence Map Modules}
The synthesis networks are prone to generating incorrect radiographic features at or near the edges, since they are high frequency components. Thus, a special attention in those regions where the network tends to be uncertain can improve the MR image synthesis task. To address this
issue, a synthesis module and a confidence map module are added on each branch of the stretch-out up-sampling block (see Synthesis Module (SM) and Confidence Map Module (CM) in Figure~\ref{fig2}). Specifically, we estimate the intermediate synthesis results at 0.5 scale size of the final output by SM and measure the confidence map which gives attentions to the 
the uncertain regions by CM. The confidence score at each pixel is a measurement of certainty about the intermediate results computed at each pixel. Confidence maps produce high confidence values (i.e
close to 1) from the regions where the network is certain about the
synthesized intensity values, and assign low confidence scores (i.e
close to 0) for those pixels where the network is uncertain. To this end,
we can block the erroneous regions by combing confidence maps and the intermediate results. Thus, the masked intermediate results is returned to the subsequent networks, which makes the network more attentive in the
uncertain regions.

As shown in Figure~\ref{fig3}, feature maps at scale $\times$0.5 ($f_{\times 0.5}$) are given as input to SM to compute the intermediate results of each MR sequence at scale $\times$0.5.
SM is a sequence of four convolutional blocks. Then, we feed the
estimated intermediate results and the feature maps as inputs
to CM for computing the confidence scores at every pixel. CM is also a sequence of four convolutional blocks. 
Finally, the confidence-masked intermediate results (i.e. the element-wise multiplication between $\hat{y}_{\times 0.5}$ and $c_{\times 0.5}$) combining with feature maps at scale $\times 0.5$ are fed back to the network to guide the subsequent layers to produce final output. Inspired by modeling the data dependent aleatoric uncertainty \cite{au44,au45}, we define the confidence map loss as follows:
\setlength{\belowdisplayskip}{0pt} \setlength{\belowdisplayshortskip}{0pt}
\setlength{\abovedisplayskip}{0pt} \setlength{\abovedisplayshortskip}{0pt}
\begin{equation} \label{eq:cf}
\begin{aligned} 
\mathcal{L}_{\text{CM}}(f_{\times 0.5}) &= c_{\times 0.5} \otimes \|\hat{y}_{\times 0.5}-y_{\times 0.5}\|_{1}-  \lambda_{\text{cm}}  C, \\
 \hat{y}_{\times 0.5} &= \text{SM}(f_{\times 0.5}),\\
 \hat{c}_{\times 0.5} &= \text{CM}(f_{\times 0.5}\oplus\hat{y}_{\times 0.5}),\\
 C &= \sum_i\sum_j  \log(c_{\times 0.5}^{ij}),
\end{aligned}
\end{equation}
where $\otimes$, $\oplus$ are the element-wise multiplication and the channel-wise concatenation, respectively. $c_{\times 0.5}^{ij}$ represents the confidence score at the $i$th row, $j$th column of the confidence map $c_{\times 0.5}$. $\hat{y}_{\times 0.5}$ is intermediate synthesis results produced by the decoder part. In $\mathcal{L}_{\text{CM}}$, the first term minimizes the L1 difference between $\hat{y}_{\times 0.5}$ and $y_{\times 0.5}$, and the values of $c_{\times 0.5}$ as well. To avoid trivial solution (i.e. $c_{\times 0.5}^{ij} = 0,   \forall i, j$), we introduce the second term as a regularizer. $\lambda_{\text{cm}}$ is a constant adjusting the weight of this regularization term $C$. Similar loss has been used for image restoration and ultrasound segmentation tasks in \cite{au50,au51}. To the best of our knowledge, our method is the first attempt to introduce this kind of loss in MR synthesis tasks.
\begin{figure}
	\centering
	\includegraphics[width=2.5in]{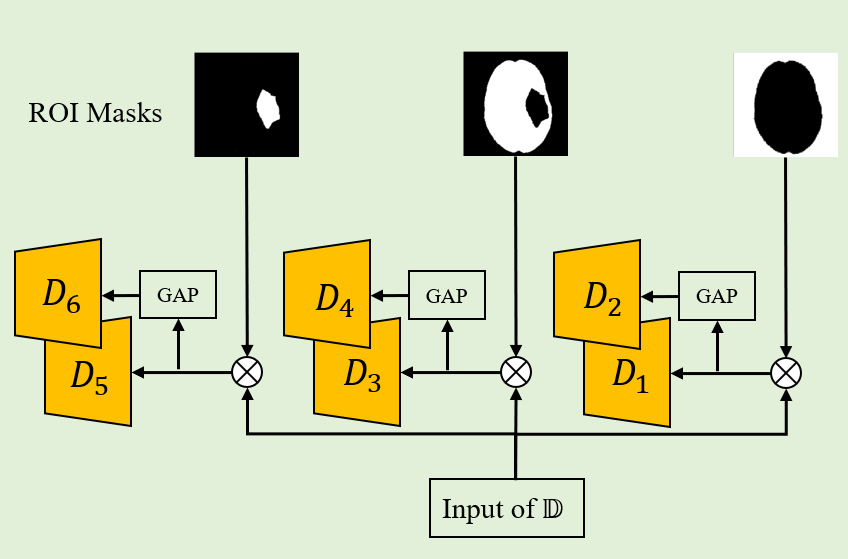}
	\caption{An overview of multi-scale label-wise discriminators. ROI masks are produced from reorganized input lesion masks. We denote $\otimes$ as the element-wise multiplication operation. GAP is the global average pooling that generates 0.5 scale size of input. $\mathbb{D}$ is a set of discriminators. }
	\label{fig:4}
\end{figure}

\subsection{Multi-scale Label-wise Discriminators}
In order to achieve large receptive field in discriminators without introducing deeper networks, we adopt multi-scale PatchGAN discriminators \cite{au8}, which have identical network architectures but take multi-scale inputs \cite{au9}. To distinguish between real and synthesized images, conventional discriminators operate on whole input. However, optimizing generator to produce realistic images in each ROI cannot be guaranteed by discriminating on holistic images, since the difficulty of synthesizing images in different regions is varying. To address this issue, we introduce label-wise discriminators. Based on the radiographic features, original lesion segmentation masks are reorganized into 3 ROIs, including background, normal brain, and lesion. As shown in Figure~\ref{fig:4}, the input of each discriminator is masked by corresponding ROI. Since the proposed discriminators are in a multi-scale setting,  for each ROI there are 2 discriminators that operate on the original and a down-sampled  $\times$0.5 scales. Thus, there are in total 6 discriminators for 3 ROIs and we refer to these set of discriminators as $\mathbb{D}=\{D_1, D_2, D_3, D_4, D_5, D_6\}$. In particular, \{$D_1$,$D_2$\},\{$D_3$,$D_4$\}, and \{$D_5$,$D_6$\} operate on the original and down-sampled versions of background, normal brain, and lesion, respectively. An overview of the proposed discriminators is given in Figure~\ref{fig:4}. The objective function corresponding to the discriminators  $\mathcal{L}_{\text{GAN}}(G,D_k)$ is as follows:
\begin{equation} \label{eq:3}
\begin{aligned} 
\mathcal{L}_{\text{GAN}}(G,D_k)= & \mathbb{E}_{(x^\prime,y^\prime)}[\log D_{k}(x^\prime,y^\prime)] \\
+ & \mathbb{E}_{x^\prime}[\log (1- D_{k}(x^\prime, G^\prime(x)))], \\
\mathcal{L}_{\text{GAN}}(G,D) = & \sum_{k=1}^{6} \mathcal{L}_{\text{GAN}} (G,D_k ),
\end{aligned}
\end{equation}
where $x$ and $y$ are paired input and real multi-modal MR images, respectively. Here, $x^\prime \triangleq m_k\otimes x$, $y^\prime \triangleq m_k\otimes y$, and $G^\prime(x) \triangleq m_k\otimes G(x)$, where  $\otimes$ denotes element-wise multiplication and $m_k$ corresponds to the ROI mask.  For simplicity, we omit the down-sampling operation in this equation.

\subsection{Training Using Paired Data}
A multi-task loss is designed to train the generator and the discriminators in an adversarial setting. Instead of only using the conventional adversarial loss $\mathcal{L}_{\text{GAN}}$, we also adopt a feature matching loss $\mathcal{L}_{\text{FM}}$ \cite{au9} to stabilize training, which optimizes generator to match these intermediate representations from the real and the synthesized images in multiple layers of the discriminators. For discriminators, $\mathcal{L}_{\text{FM}}(G,D_k)$ is defined as follows:  
\begin{equation} \label{eq:4}
\begin{aligned}
\mathcal{L}_{\text{FM}}(G,D_k)= & \sum_{i}^{T}  \frac{1}{N_{i}} \left[\|D_{k}^{(i)}(x^\prime,y^\prime)- D_{k}^{(i)}(x^\prime, G^\prime(x)\|_{2}^{2}\right] \\
\mathcal{L}_{\text{FM}}(G,D) =& \sum_{k=1}^{6}\mathcal{L}_{\text{FM}}(G,D_k ),
\end{aligned}
\end{equation}
where $D_{k}^{(i)} $ denotes the $i$th layer of the discriminator $D_{k}$, $T$ is the total number of layers in $D_{k}$ and $N_i$ is the number of elements  in the $i$th layer. If we perform lesion segmentation on images, it is worth to note that there is a consistent relation between the prediction and the real one serving as input for the generator. Lesion labels are usually occluded with each other and brain anatomic structure, which causes ambiguity for synthesizing realistic MR images. To tackle this problem, we propose a lesion shape consistency loss $\mathcal{L}_{\text{SC}}$ by adding a  U-net \cite{au11} segmentation module (see Figure~\ref{fig2}) that regularizes the generator to obey this consistency relation. We adopt Generalized Dice Loss (GDL) \cite{au16} to measure the difference between the predicted and real segmentation maps and is defined as follows
\begin{equation} \label{eq:7}
\text{GDL}(R,S) =  1 - \frac{2\sum_i^N r_i s_i }{\sum_i^N r_i + \sum_i^N s_i },
\end{equation}
where $R$ denotes the ground truth and $S$ is the segmentation result. $r_i$ and $s_i$ represent the ground truth and predicted probability maps  at each pixel $i$, respectively. $N$ is the total number of pixels. The lesion shape consistency loss $\mathcal{L}_{\text{SC}}$ is then defined as follows      
\begin{equation} \label{eq:5}
\begin{aligned}
\mathcal{L}_{\text{SC}}(U) = \text{GDL}(s,U(y)) + \text{GDL}(s,U(G(x))),
\end{aligned}
\end{equation}
where $U(y)$ and $U(G(x))$ represent the predicted lesion segmentation probability maps by taking $y$ and $G(x)$ as inputs in the segmentation module, respectively. $s$ denotes the ground truth lesion segmentation map. The final multi-task objective function for training CG-SAMR is defined as
\begin{equation} \label{eq:6}
\begin{aligned}
\min\limits_{\text{G,U}}(\max\limits_{\text{D}} \mathcal{L}_{\text{GAN}} (G,D))  + & \lambda_{1} \mathcal{L}_{\text{FM}}(G,D) \\
+& \lambda_{2}\mathcal{L}_{\text{SC}}(U) + \lambda_{3}\mathcal{L}_{\text{CM}}(f_{\times 0.5}),
\end{aligned}
\end{equation}
where $\lambda_{1}$,  $\lambda_{2}$ and $\lambda_{3}$ three parameters that control the importance of each loss.

\subsection{Training Using Unpaired Data}
\begin{figure}
	\centering
	\includegraphics[width=.6\columnwidth]{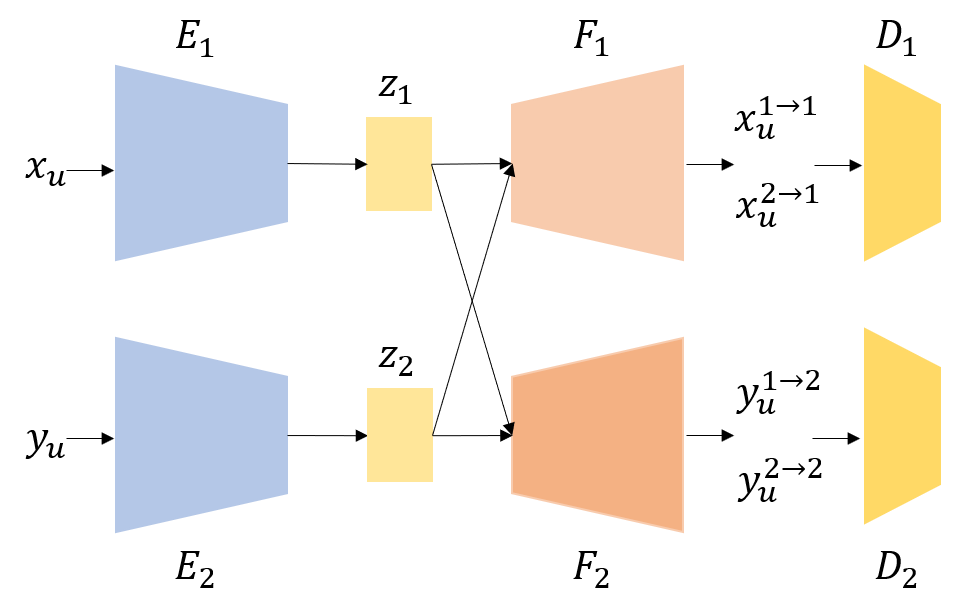}
\vskip -10pt	\caption{The schematic of the proposed method corresponding to  training using unpaired data. $E_1$ and $E_2$ are two encoders mapping input to the latent codes. $F_1$ is a decoder with symmetric architecture as encoders mapping the latent codes to domain 1. $F_2$ is a decoder that is used in CG-SAMR mapping the latent codes to multi-modal MR images (domain2). $D_1$ and $D_2$ are two discriminators for domain 1 and domain 2.  \label{fig6}}
\end{figure}

\begin{table*}[]
	\setlength{\tabcolsep}{0.6pt}
	\centering
	\caption{Quantitative comparison. Quality of the synthesized data under paired data training is measured by pixel accuracy. Lesion indicates the union of edema, cavity, and tumor. Brain represent the holistic brain region. Here, the unit is in percent (\%).}\label{tab1}
	\scriptsize
	\begin{tabular}{cccccccccccccccccccccccccc}
		\hline
		& \multicolumn{5}{c}{Pix2Pix \cite{au8}}                                  & \multicolumn{5}{c}{Pix2PixHD \cite{au9}}                                & \multicolumn{5}{c}{Shin et al. \cite{au14}}                              & \multicolumn{5}{c}{SAMR \cite{au59}}                                  & \multicolumn{5}{c}{CG-SAMR (our)}                                             \\ \hline
		& Edema & Cavity & Tumor & Lesion & \multicolumn{1}{c|}{Brain} & Edema & Cavity & Tumor & Lesion & \multicolumn{1}{c|}{Brain} & Edema & Cavity & Tumor & Lesion & \multicolumn{1}{c|}{Brain} & Edema & Cavity & Tumor & Lesion & \multicolumn{1}{c|}{Brain} & Edema         & Cavity        & Tumor         & Lesion        & Brain         \\ \hline
		\multicolumn{1}{c|}{$APT$w}   & 50.8  & 42.1   & 48.2  & 48.8   & \multicolumn{1}{c|}{51.0}  & 55.0  & 42.1   & 51.2  & 51.5   & \multicolumn{1}{c|}{52.9}  & 45.2  & 40.0   & 42.0  & 43.9   & \multicolumn{1}{c|}{46.8}  & 65.9  & 52.7   & 63.1  & 63.8   & \multicolumn{1}{c|}{55.1}  & 67.1          & 51.3          & 64.3          & 64.2          & 56.1          \\
		\multicolumn{1}{c|}{$T_1$w}    & 54.6  & 56.2   & 49.2  & 53.5   & \multicolumn{1}{c|}{42.4}  & 54.0  & 53.0   & 47.8  & 52.7   & \multicolumn{1}{c|}{44.2}  & 72.6  & 71.7   & 68.0  & 71.8   & \multicolumn{1}{c|}{73.9}  & 73.0  & 69.0   & 67.5  & 72.8   & \multicolumn{1}{c|}{53.4}  & 76.0          & 67.8          & 71.1          & 75.0          & 57.4          \\
		\multicolumn{1}{c|}{$FLAIR$} & 51.7  & 41.0   & 44.7  & 48.5   & \multicolumn{1}{c|}{57.7}  & 47.1  & 36.3   & 46.3  & 44.5   & \multicolumn{1}{c|}{58.8}  & 60.1  & 41.9   & 51.9  & 56.9   & \multicolumn{1}{c|}{65.8}  & 75.4  & 61.5   & 68.1  & 73.1   & \multicolumn{1}{c|}{68.1}  & 78.2          & 67.4          & 71.5          & 76.4          & 71.6          \\
		\multicolumn{1}{c|}{$T_2$w}    & 52.1  & 52.3   & 42.5  & 51.2   & \multicolumn{1}{c|}{57.3}  & 50.6  & 59.3   & 46.4  & 50.3   & \multicolumn{1}{c|}{57.8}  & 65.6  & 55.5   & 56.3  & 63.1   & \multicolumn{1}{c|}{70.0}  & 76.7  & 77.7   & 71.2  & 77.3   & \multicolumn{1}{c|}{68.9}  & 81.0          & 77.7          & 74.3          & 80.7          & 72.5          \\
		\multicolumn{1}{c|}{Gd-$T_1$w}   & 70.4  & 57.7   & 38.1  & 63.3   & \multicolumn{1}{c|}{58.5}  & 72.3  & 58.5   & 37.4  & 65.0   & \multicolumn{1}{c|}{60.5}  & 74.4  & 64.8   & 38.7  & 67.5   & \multicolumn{1}{c|}{71.4}  & 81.2  & 67.7   & 64.2  & 78.0   & \multicolumn{1}{c|}{69.9}  & 83.1          & 69.3          & 62.6          & 79.1          & 73.2          \\ \hline
		\multicolumn{1}{c|}{Avg.}  & 55.9  & 49.9   & 44.5  & 53.1   & \multicolumn{1}{c|}{53.4}  & 55.8  & 49.8   & 45.8  & 52.8   & \multicolumn{1}{c|}{54.8}  & 63.6  & 54.8   & 51.4  & 60.6   & \multicolumn{1}{c|}{65.6}  & 74.4  & 65.7   & 66.8  & 73.0   & \multicolumn{1}{c|}{63.1}  & \textbf{77.1} & \textbf{66.7} & \textbf{68.8} & \textbf{75.1} & \textbf{66.2} \\ \hline
	\end{tabular}
\end{table*}

Figure~\ref{fig6} shows the schematic of the proposed method corresponding to  training using unpaired data. Our framework is based on the proposed CG-SAMR network and an additional GAN: $\text{GAN}_1 = \{E_1,F_1,D_1\}$ and $\text{GAN}_2 = \{E_2,F_2,D_2\}$. Denote the set of lesion segmentation maps and anatomic prior as domain 1 and the set of multi-modal MR images as domain 2. Here, we denote \textbf{unpaired} instances in domain 1 and 2 as $x_u$ and $y_u$, respectively. In $\text{GAN}_1$, $D_1$ aims to evaluate whether the translated unpaired instances are realistic. It outputs true for real instances sampled from the domain 1 and false for instances generated by $F_1$. As shown in Figure~\ref{fig6}, $F_1$ can generate two types of instances: (1) instances from the reconstruction stream $x_u^{1\rightarrow 1} = F_1(E_1(x_u))$, and (2) instances from the cross-domain stream $x_u^{2\rightarrow 1} = F_1(E_2(y_u))$. We have similar properties in $\text{GAN}_2$, but the decoder $F_2$ is replaced by the corresponding decoder part in CG-SAMR. Thus, we can realize confidence-guided customized synthesis for each MR sequence under unpaired data training. The objective functions for reconstruction streams are defined as follows
\begin{equation} \label{eq:recon}
\begin{aligned}
\mathcal{L}_{\text{recon}_1} = & \|x_u-x_u^{1\rightarrow 1}\|_1, \\
\mathcal{L}_{\text{recon}_2} = & \|y_u-y_u^{2\rightarrow 2}\|_1 + \mathcal{L}_{\text{CM}}(f_{\times 0.5}|y_u),
\end{aligned}
\end{equation}
where $\mathcal{L}_{\text{CM}}$ is defined in equation~(\ref{eq:cf}) and $F_2$ is a decoder network with the same architecture as used in CG-SAMR. We denote the feature maps used for $\mathcal{L}_{\text{CM}}$ in $F_2$ as $f_{\times 0.5}|y_u$ when decoding the latent code obtained by encoding $y_u$. The objective functions of cross-domain streams can be expressed as follows
\begin{equation} \label{eq:cross}
\begin{aligned}
\mathcal{L}_{\text{GAN}_1} = & \mathbb{E}_{(x_u)}[\log D(x_u)]
+  \mathbb{E}_{(z_2)}[\log (1- D_{1}(F_1(z_2)))],\\
\mathcal{L}_{\text{GAN}_2} = & \mathbb{E}_{(y_u)}[\log D(y_u)] 
+  \mathbb{E}_{(z_1)}[\log (1- D_{2}(F_2(z_1)))], \\
\end{aligned}
\end{equation}
where $z_1$ and $z_2$ are the latent codes, $z_1 = E_1(x_u)$, $z_2 = E_2(y_u).$ 
Simply relaying on the reconstruction stream and adversarial training (i.e. cross-domain streams) cannot guarantee to learn the desired mapping function.  To reduce the number of possible mapping functions, we require the learned mapping functions to obey cycle-consistent constraint (i.e. $x_u \rightarrow y_u^{1\rightarrow2} \rightarrow F_1(E_2(y_u^{1\rightarrow2})) \approx x_u )$ \cite{au39}. The objective functions for cycle-reconstruction streams are defined as follows 
\begin{equation} \label{eq:cyc}
\begin{aligned}
\mathcal{L}_{\text{cyc}_1} = & \|x_u-F_1(E_2(y_u^{1\rightarrow 2}))\|_1, \\
\mathcal{L}_{\text{cyc}_2} = & \|y_u-F_2(E_1(x_u^{2\rightarrow 1}))\|_1 + \mathcal{L}_{\text{CM}}(f_{\times 0.5}|x_u^{2\rightarrow 1}).
\end{aligned}
\end{equation}
The overall objective function used to train the UCG-SAMR in unsupervised setting is defined as follows 
\begin{equation} \label{eq:unpair}
\begin{aligned}
G^* = &\min\limits_{\{E_1,F_1,E_2,F_2\}}\max\limits_{\{D_1,D_2\}} \mathcal{L}_{\text{domain}_1} + \mathcal{L}_{\text{domain}_2}, \;\text{where}\\
\mathcal{L}_{\text{domain}_1} = &  \mathcal{L}_{\text{recon}_1} +  \mathcal{L}_{\text{GAN}_1}
+  \mathcal{L}_{\text{cyc}_1},\;\;\text{and} \\
\mathcal{L}_{\text{domain}_2} = &  \mathcal{L}_{\text{recon}_2} +  \mathcal{L}_{\text{GAN}_2}
+  \mathcal{L}_{\text{cyc}_2}.
\end{aligned}
\end{equation}

\begin{figure}
	\centering
	\includegraphics[width=\columnwidth]{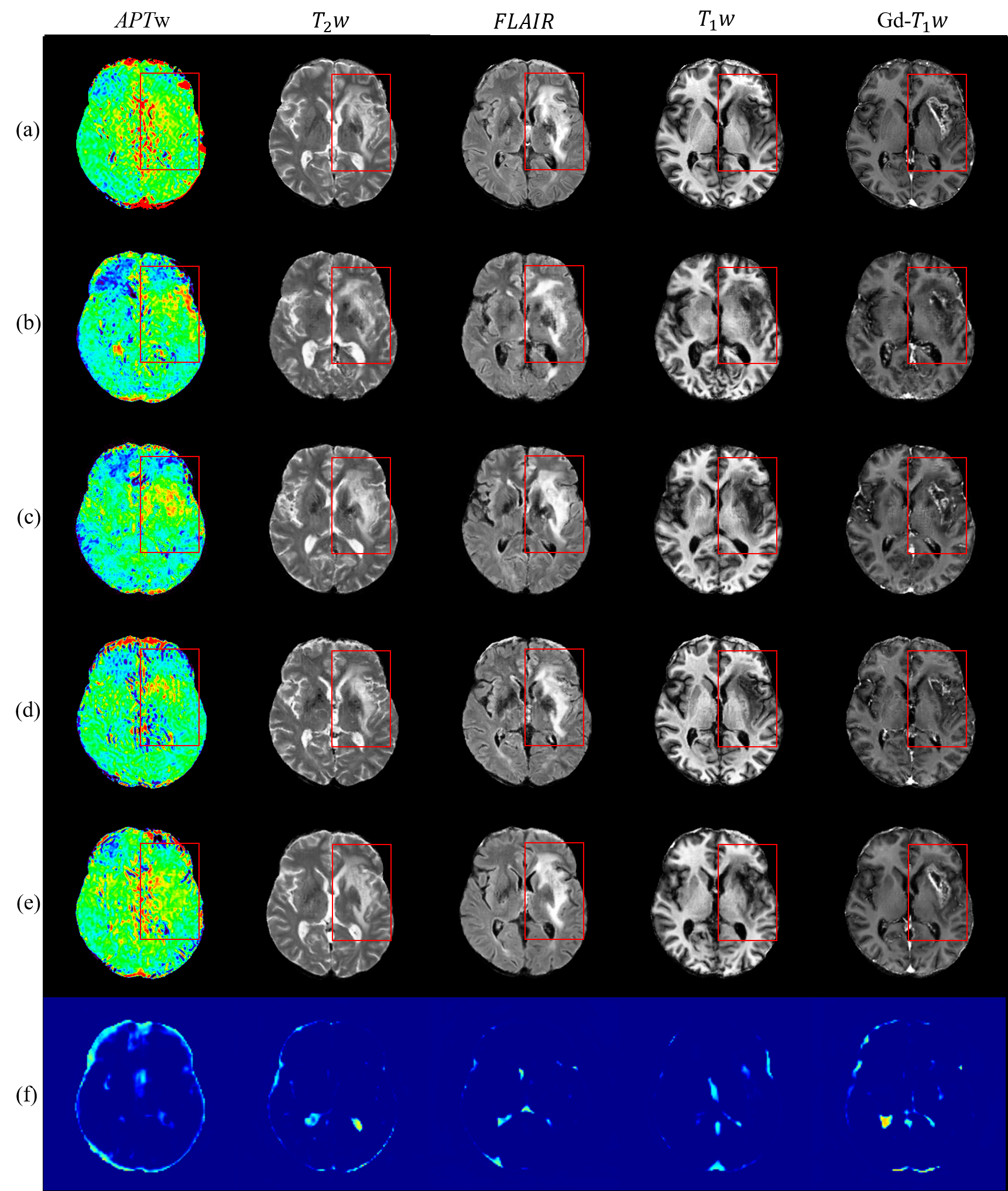}
\vskip -8pt	\caption{Qualitative comparison of different methods under paired data training.  The same lesion mask is used to synthesize images from different methods. (a) Real data (ground truth). (b) Pix2Pix \cite{au8}. (c) Pix2PixHD\cite{au9}. (d) Shin et al. \cite{au14}. (e) CG-SAMR (our). (f) Confidence maps from CG-SAMR. Red boxes indicate the lesion region. \label{fig7}}
\end{figure}

\begin{table}[ht]
	\setlength{\tabcolsep}{4.0pt}
	\centering
	\caption{Quantitative results corresponding to image segmentation when  the synthesized data is used for data augmentation. For each experiment, the first row reports the percentage of synthesized/real data for training and the number of instances of synthesized/real data in parentheses. Exp.3 reports the results of baseline trained only by real data.}\label{tab2}
	\scriptsize
	\begin{tabular}{ccccccc}
		\hline
		\multicolumn{7}{c}{Exp.1:  50\% Synthesized+ 50\% Real (1080 + 1080)}                                                                                                 \\ \hline
		\multicolumn{1}{l}{}                       & \multicolumn{3}{c}{Dice Score}                                        & \multicolumn{3}{c}{Hausdorff95 Distance}         \\ \hline
		\multicolumn{1}{c|}{}                      & Edema          & Cavity         & \multicolumn{1}{c|}{Tumor}          & Edema          & Cavity         & Tumor          \\
		\multicolumn{1}{c|}{Pix2Pix \cite{au8}}      & 0.589          & 0.459          & \multicolumn{1}{c|}{0.562}          & 13.180         & 21.003         & 10.139         \\
		\multicolumn{1}{c|}{Pix2PixHD \cite{au9}}    & 0.599          & 0.527          & \multicolumn{1}{c|}{0.571}          & 17.406         & 8.606          & 10.369         \\
		\multicolumn{1}{c|}{Shin et al. \cite{au14}} & 0.731          & 0.688          & \multicolumn{1}{c|}{0.772}          & 7.306          & 6.290          & 6.294          \\
		\multicolumn{1}{c|}{SAMR \cite{au59}}                  & 0.794          & 0.813          & \multicolumn{1}{c|}{0.821}          & 6.049          & 1.568          & 2.293          \\
		\multicolumn{1}{c|}{CG-SAMR (our)}         & \textbf{0.804} & \textbf{0.839} & \multicolumn{1}{c|}{\textbf{0.828}} & \textbf{4.166} & \textbf{1.381} & \textbf{1.810} \\ \hline
		\multicolumn{7}{c}{Exp.2:  25\% Synthesized+ 75\% Real (540 + 1080)}                                                                                                  \\ \hline
		\multicolumn{1}{c|}{Pix2Pix \cite{au8}}      & 0.602          & 0.502          & \multicolumn{1}{c|}{0.569}          & 10.706         & 9.431          & 10.147         \\
		\multicolumn{1}{c|}{Pix2PixHD \cite{au9}}    & 0.634          & 0.514          & \multicolumn{1}{c|}{0.663}          & 17.754         & 9.512          & 9.061          \\
		\multicolumn{1}{c|}{Shin et al. \cite{au14}} & 0.673          & 0.643          & \multicolumn{1}{c|}{0.708}          & 14.835         & 7.798          & 6.688          \\
		\multicolumn{1}{c|}{SAMR \cite{au59}}                  & 0.745          & 0.780          & \multicolumn{1}{c|}{0.772}          & 8.779          & 6.757          & 4.735          \\
		\multicolumn{1}{c|}{CG-SAMR (our)}         & \textbf{0.756} & \textbf{0.793} & \multicolumn{1}{c|}{\textbf{0.773}} & \textbf{7.676} & \textbf{6.258} & \textbf{4.325} \\ \hline
		\multicolumn{7}{c}{Exp.3:  0\% Synthesized + 100\% Real (0 + 1080)}                                                                                                   \\ \hline
		\multicolumn{1}{c|}{Baseline}              & \textbf{0.646} & \textbf{0.613} & \multicolumn{1}{c|}{\textbf{0.673}} & \textbf{8.816} & \textbf{7.856} & \textbf{7.078} \\ \hline
	\end{tabular}
\end{table}

\begin{table*}[]
	\setlength{\tabcolsep}{3.0pt}
	\centering
	\caption{Quantitative comparison.  The quality of synthesized data under unpaired data training is measured by pixel accuracy. Lesion indicates the union of edema, cavity, and tumor. Brain represent the holistic brain region.}\label{tab3}
	\scriptsize
	\begin{tabular}{cccccccccccccccc}
		\hline
		& \multicolumn{5}{c}{CycleGAN~\cite{au39}}                                 & \multicolumn{5}{c}{UNIT~\cite{au48}}                                     & \multicolumn{5}{c}{UCG-SAMR (our)}                 \\ \hline
		& Edema & Cavity & Tumor & Lesion & \multicolumn{1}{c|}{Brain} & Edema & Cavity & Tumor & Lesion & \multicolumn{1}{c|}{Brain} & Edema & Cavity & Tumor & Lesion & Brain \\ \hline
		\multicolumn{1}{c|}{$APT$w}   & 51.3  & 32.8   & 39.9  & 47.3   & \multicolumn{1}{c|}{47.3}  & 44.1  & 33.7   & 41.3  & 42.2   & \multicolumn{1}{c|}{42.3}  & 51.5  & 36.8   & 44.7  & 48.2   & 43.0  \\
		\multicolumn{1}{c|}{$T_1$w}    & 35.5  & 23.6   & 34.2  & 34.9   & \multicolumn{1}{c|}{53.2}  & 64.5  & 65.1   & 64.4  & 63.1   & \multicolumn{1}{c|}{68.5}  & 66.4  & 60.5   & 68.7  & 65.2   & 68.0  \\
		\multicolumn{1}{c|}{$FLAIR$} & 56.8  & 33.2   & 35.2  & 49.2   & \multicolumn{1}{c|}{60.1}  & 55.3  & 37.9   & 49.9  & 52.3   & \multicolumn{1}{c|}{62.5}  & 65.6  & 37.1   & 58.0  & 60.4   & 65.9  \\
		\multicolumn{1}{c|}{$T_2$w}    & 67.0  & 6.1    & 54.7  & 57.1   & \multicolumn{1}{c|}{64.0}  & 63.7  & 41.4   & 52.0  & 58.8   & \multicolumn{1}{c|}{66.6}  & 69.1  & 48.4   & 59.5  & 65.7   & 68.5  \\
		\multicolumn{1}{c|}{$Gd$-$T_1$w}   & 47.5  & 45.1   & 22.2  & 42.5   & \multicolumn{1}{c|}{62.4}  & 65.7  & 65.3   & 42.2  & 62.3   & \multicolumn{1}{c|}{69.7}  & 72.5  & 65.8   & 45.3  & 67.8   & 69.9  \\ \hline
		\multicolumn{1}{c|}{Avg.}  & 51.6  & 28.2   & 37.2  & 46.2   & \multicolumn{1}{c|}{57.4}  & 58.7  & 48.7   & 50.0  & 55.7   & \multicolumn{1}{c|}{61.9}  & \textbf{65.0}  & \textbf{49.7}   & \textbf{55.2}  & \textbf{61.5}   & \textbf{63.1}  \\ \hline
	\end{tabular}
\end{table*}

\begin{table}[]
	\setlength{\tabcolsep}{3.0pt}
	\centering
	\caption{Quantitative evaluation of the segmentation performance of different method under unpaired data training. }\label{tab4}
	\scriptsize
	\begin{tabular}{cccc|ccc}
		\hline
		& \multicolumn{3}{c|}{Dice Score} & \multicolumn{3}{c}{Hausdorff95 Distance} \\ \hline
		\multicolumn{1}{c|}{}         & Edema    & Cavity    & Tumor    & Edema        & Cavity      & Tumor       \\ \cline{2-7}
		\multicolumn{1}{c|}{CycleGAN~\cite{au39}} & 0.333    & 0.01      & 0.073    & 18.647       & 30.859      & 39.611      \\
		\multicolumn{1}{c|}{UNIT~\cite{au48}}     & 0.527    & 0.368     & 0.506    & 9.008        & 11.225      & 10.183      \\
		\multicolumn{1}{c|}{UCG-SAMR (our)}      & \textbf{0.558}    & \textbf{0.393}     & \textbf{0.613}    & \textbf{8.321}        & \textbf{11.130}      & \textbf{7.044}   \\ \hline   
	\end{tabular}
\end{table}
\section{Experiments and Results}
In this section, we first discuss the data acquisition and training details. Then, the experimental setup, evaluations of the proposed synthesis methods against a set of recent state-of-the-art approaches, and comprehensive ablation studies are presented.  
\label{sec4}
\subsection{Data Acquisition}
\label{sec:da}
This study was approved by the Institutional Review Board (IRB) and conducted in accordance with the U.S. Common Rule, and consent form was waved. Patients enrollment criteria are: at least 20 years old; initial diagnosis of pathologically proven primary malignant glioma; status post initial surgery and chemoradiation. There are 90 patients who are involved in this study. MRI scans were acquired by a 3T human MRI scanner (Achieva; Philips Medical Systems) by using a body coil excite and a 32-channel phased-array coil for reception \cite{au5}. $T_1$w, Gd-$T_1$w, $T_2$w, $FLAIR$, and $APT$w MRI sequences were collected for each patient. Imaging parameters for $APT$w can be summarized as: field of view (FOV), 212 $\times$  212 $\times$ 66 $mm^{3}$; resolution, 0.82 $\times$  0.82 $\times$  4.4  $mm^3$; size of matrix, 256 $\times$  256 $\times$ 15. Other anatomic MRI sequences were  acquired with Imaging parameters: FOV, 212 $\times$  212 $\times$ 165 $mm^{3}$; resolution, 0.41 $\times$  0.41 $\times$  1.1  $mm^3$; size of matrix, 512 $\times$  512 $\times$ 150. Co-registration between $APT$w and anatomic sequences \cite{au17}, skull stripping \cite{au20}, N4-bias field correction \cite{au18}, and MRI standardization \cite{au19} were performed sequentially. After preprocessing, the final volume size of each sequence is 256 $\times$  256 $\times$ 15. For every collected volume, lesion were manually annotated by an expert neuroradiologist into three labels: edema, cavity and tumor. Then, a multivariate template construction tool \cite{au21} was used to create the group average for each sequence (atlas). 1350 instances with the size of 256 $\times$ 256 $\times$  5 were extracted from volumetric data, where 5 corresponds to five MRI sequences. For every instance, the one corresponding atlas slice and two adjunct (in axial direction) atlas slices were extracted to provide the prior of human brain anatomy in paired data training. The WM, GM, CSF probability masks were also extracted to provide anatomic prior used in the unsupervised case by SPM12 \cite{au52}. We split these instances randomly into 1080 (80\%) for training and 270 (20\%)  for testing. Since the data was split on the patient level, training and testing data did not include the instances from the same patient.

\subsection{Implementation Detail}
 The CG-SAMR synthesis model was trained based on the final objective function equation~(\ref{eq:6}) using the Adam optimizer \cite{au21}. $\lambda_{1}$, $\lambda_{2}$ and $\lambda_{3} $  were set equal to 5, 1 and 1, respectively. Hyperparameters are set as follows: constant learning rate of 2 $\times 10^{-4}$ for the first 250 epochs then linearly decaying to 0; 500 maximum epochs; batch size of 8. $\lambda_{\text{cm}}$ in equation~(\ref{eq:cf}) initially was set equal to 0.1. When the mean of scores in confidence maps $c_{\times 0.5}$ is greater than 0.7, $\lambda_{\text{cm}}$ was set equal to 0.03. Hyperparameters for unpaired data training are set as follows: constant learning rate of 2 $\times 10^{-4}$ for the first 400 epochs then linearly decaying to 0; 800 maximum epochs; batch size of 1. To further evaluating the effectiveness of the synthesized MRI sequences on data augmentation, we leveraged U-net \cite{au11} to train lesion segmentation models. U-net \cite{au11} was trained by the Adam optimizer \cite{au21}. Hyperparameters are set as follows: constant learning rate of 2 $\times 10^{-4}$ for the first 100 epochs then linearly decaying to 0; 200 maximum epochs; batch size of 16. In the segmentation training, all the synthesized data was produced from randomly manipulated lesion masks by CG-SAMR. For evaluation, we always keep 20\%  of the data unseen for both of the synthesis and segmentation models.

\begin{figure}
	\centering
	\includegraphics[width=.9\columnwidth]{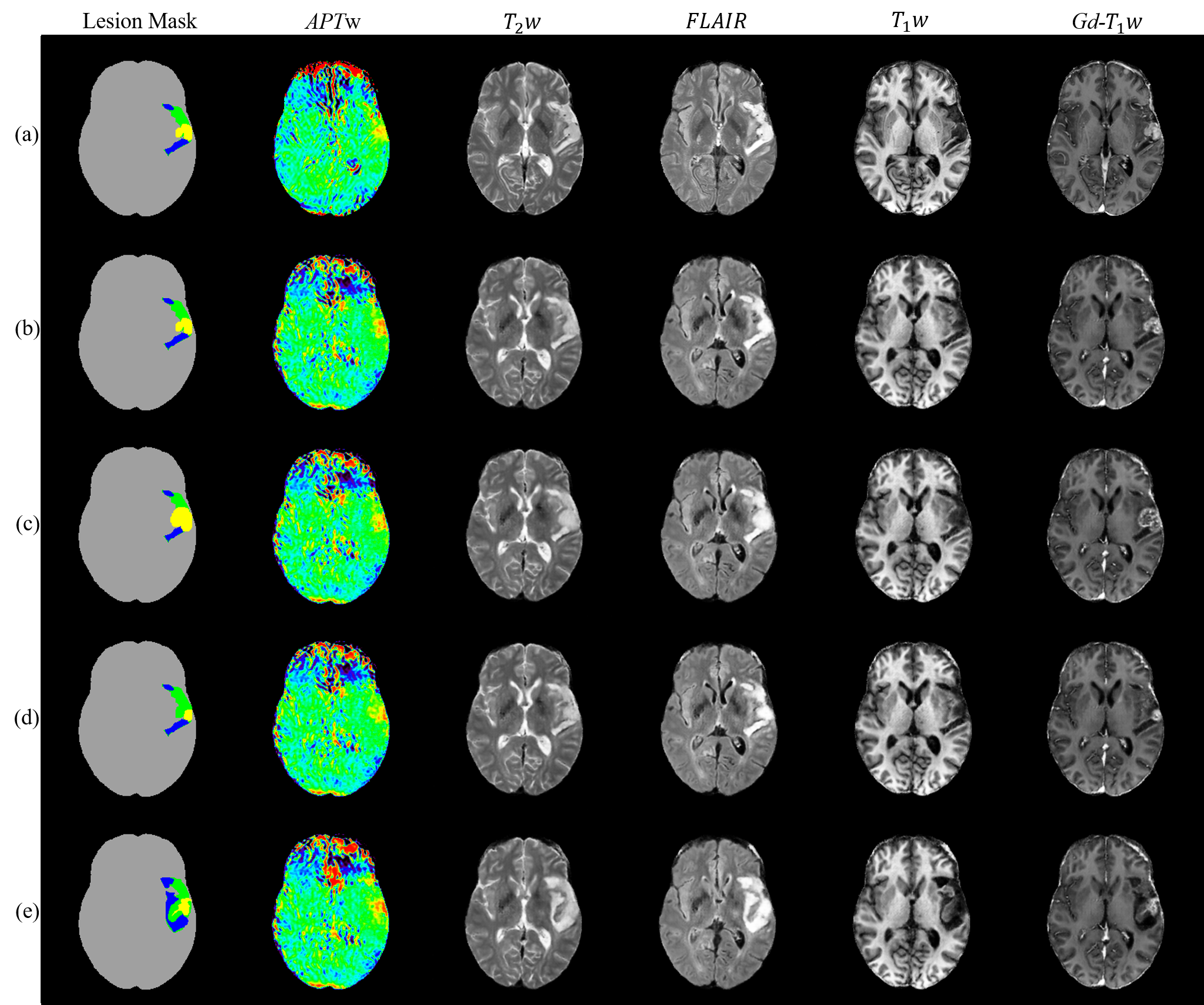}
	\caption{Examples of lesion mask manipulations in CG-SAMR. (a) Real images (ground truth). (b) Synthesized images from the original mask. (c) Synthesized images by increasing tumor size to 100\%. (d) Synthesized images by shrinking tumor size to 50\%. (e) Synthesized images by replacing lesion from another slice. In lesion masks, gray, green, yellow, and blue represent normal brain, edema, tumor, and cavity, respectively. \label{fig8}}
\end{figure}

\begin{figure}
	\centering
	\includegraphics[width=.9\columnwidth]{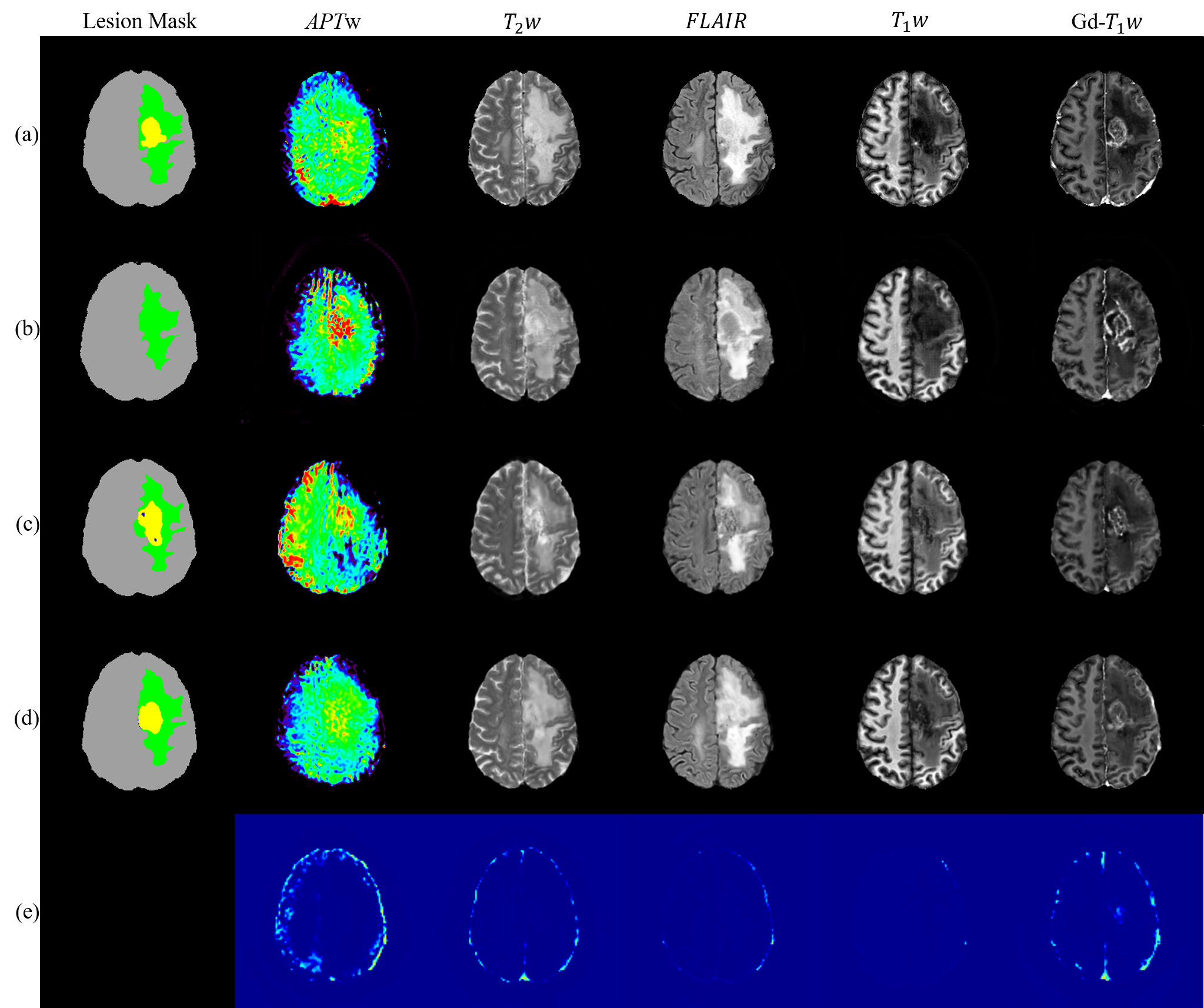}
	\caption{Qualitative comparison of segmentation and synthesis performance under unpaired data training. (a) Real data (ground truth). (b) CycleGAN \cite{au39}. (c) UNIT \cite{au9}. (d) UCG-SAMR (our). (e) Confidence maps from UCG-SAMR.  In lesion masks, gray, green, yellow, and blue represent normal brain, edema, tumor, and cavity, respectively. \label{fig9}}
\end{figure}

\subsection{Results Corresponding to Supervised Training} We evaluate the performance of our method against the following recent state-of-the-art generic synthesis methods: Pix2Pix \cite{au8}, Pix2PixHD \cite{au9} as well as MRI synthesis methods: Shin et al. \cite{au14}, and SAMR \cite{au59}. We use pixel accuracy to compare the performance of different methods \cite{au8,au9,au39}. In particular, we calculate the difference between the synthesized data and the corresponding ground truth data and a pixel translation was counted correct if the difference was within 16 of the ground truth intensity value. Table~\ref{tab1} shows the quantitative
performance of different methods in terms of pixel accuracy. As it can be seen from this
table, our method clearly outperforms the present state-of-the-art synthesis algorithms. CG-SAMR gains improvement especially at lesion regions. Figure~\ref{fig7} presents the qualitative comparisons of the synthesized multi-modal MRI sequences from four different methods. It can be observed that Pix2Pix \cite{au8} and Pix2PixHD \cite{au9} fail to synthesize realistic looking human brain MR images. There is either an unreasonable brain ventricle or wrong radiographic features in the lesion region (see Figure~\ref{fig7} (b)(c)). Shin et al. \cite{au14} can produce realistic brain anatomic structures for anatomic MRI sequences. However, there is an obvious disparity between the synthesized and real $APT$w sequence in both normal brain and lesion region. The boundary of the synthezied lesion is also blurry  (see red boxes in see Figure~\ref{fig7} (d)). The proposed method produces more accurate radiographic features of lesions and more diverse anatomic structure based on the human anatomy prior provided by atlas.

To further evaluate the quality of the synthesized MR images, we perform data augmentation by using the synthesized images in training and then perform lesion segmentation. Evaluation metrics in BraTS challenge \cite{au3} (i.e. Dice score, Hausdorff distance (95\%)) are used to measure the performance of different methods. The data augmentation by synthesis is evaluated by the improvement for lesion segmentation models. We arbitrarily control lesion information to synthesize different number of data for augmentation. 
To simulate the piratical usage of data augmentation, we conduct experiments in the manner of utilizing all real data. In each experiment, we vary the percentage of the synthesized data to observe the contribution for data augmentation. Table~\ref{tab2} shows the calculated segmentation performance. Comparing with the baseline experiment that only uses real data, the synthesized data from pix2pix \cite{au8} and pix2pixHD  \cite{au9} degrade the segmentation performance. The performance is improved when the synthesized data from of Shin et al. \cite{au14} and SAMR \cite{au59} are used for segmentation but the proposed method outperforms the other methods by a large margin. Figure~\ref{fig8} demonstrates the robustness of the proposed model under different lesion mask manipulations (e.g. changing the size of tumor and even reassembling lesion information between lesion masks). As can be seen from this figure, our method is robust to various lesion mask manipulations.

\subsection{Results Corresponding to Unsupervised Training} 
We denote the proposed method under unpaired data training as UCG-SAMR and evaluate its performance against the following recent state-of-the-art  unsupervised synthesis methods: CycleGAN \cite{au39} and UNIT \cite{au48}. Table~\ref{tab3} shows the quantitative
synthesis performance of different methods in term of pixel accuracy. As it can be seen from this
table, our method outperforms the other state-of-the-art synthesis algorithms. On average, UCG-SAMR gains 5.8\% and 15.3\% improvement at lesion regions compared to CycleGAN \cite{au39} and UNIT \cite{au48}, respectively. Table~\ref{tab4} shows the comparison of segmentation performance for different methods. We can observe that UCG-SAMR reaches the performance upper bound (i.e.  supervised training by real paired data in Table~\ref{tab1} Exp.3). Figure.~\ref{fig9} presents the qualitative comparison of the segmentation and multi-modal MRI synthesis. It can be observed that CycleGAN \cite{au39} and UNIT \cite{au48} fail to synthesize realistic looking lesions, especially in the $APT$w and Gd-$T_1$w sequences. The proposed method produces more accurate radiographic features for each type of lesion label in both molecular and anatomic sequences. Facilitated by high-quality synthesis, the segmentation network works better than the other models as can be seen from Table~\ref{tab4}.

\subsection{Ablation Study}  
\begin{table}[]
	\setlength{\tabcolsep}{2.0pt}
	\centering
	\caption{Ablation study of designed modules in data augmentation by synthesis.}\label{tab5}
	\scriptsize
	\begin{tabular}{lcccccccc}
		\hline
		\multicolumn{1}{l}{}                      & \multicolumn{1}{l}{}  & \multicolumn{3}{c}{Dice Score}                                                              & \multicolumn{1}{l}{} & \multicolumn{3}{c}{Hausdorff95 Distance}                                                                                                      \\ \hline
		& \multicolumn{1}{c|}{} & Edema                     & Cavity                    & \multicolumn{1}{c|}{Tumor}          &                      & Edema                     & Cavity                    & \multicolumn{1}{c}{Tumor}  \\ \cline{3-9}
		w/o  Stretch-out                          & \multicolumn{1}{c|}{} & 0.677                     & 0.697                     & \multicolumn{1}{c|}{0.679}          &                      & 13.909                    & 11.481                    & \multicolumn{1}{c}{7.123}          \\
		w/o  Multi-label D                         & \multicolumn{1}{c|}{} & 0.753                     & 0.797                     & \multicolumn{1}{c|}{0.785}          &                      & 7.844                     & 2.570                     & \multicolumn{1}{c}{2.719}         \\
		
		w/o  Atlas                                & \multicolumn{1}{c|}{} & 0.684                     & 0.713                     & \multicolumn{1}{c|}{0.705}          &                      & 6.592                     & 5.059                     & \multicolumn{1}{c}{4.002}         \\
		
		w/o  $\mathcal{L}_{\text{SC}}$ & \multicolumn{1}{c|}{} & 0.728 & 0.795 & \multicolumn{1}{c|}{0.771}      &     & 8.604 & 3.024 & \multicolumn{1}{c}{3.233}          \\
		
		w/o $\mathcal{L}_{\text{CM}}$                                  & \multicolumn{1}{c|}{} & 0.794            & 0.813            & \multicolumn{1}{c|}{0.821} &      \textbf{}      & 6.049           & 1.568           & \multicolumn{1}{c}{2.293}  \\
		CG-SAMR (proposed)                                       & \multicolumn{1}{c|}{} & \textbf{0.828}            & \textbf{0.839}            & \multicolumn{1}{c|}{\textbf{0.828}} & \textbf{}            & \textbf{4.166}            & \textbf{1.381}            & \multicolumn{1}{c}{\textbf{1.810}}  \\
		\hline
	\end{tabular}
\end{table}

\begin{table}[]
	\setlength{\tabcolsep}{3.0pt}
	\centering
	\caption{Ablation study of designed modules in term of synthesis quality. The reported value is pixel accuracy in the lesion region as percent (\%).}\label{tab6}
	\scriptsize
	\begin{tabular}{lccccc|c}
		\hline
		& $APT$w  & $T_1$w   & $FLAIR$ & $T_2$w   & Gd-$T_1$w  & Avg. \\ \hline
		w/o Stretch-out  & 62.5 & 66.1 & 66.2  & 70.5 & 72.4 & 67.5 \\
		w/o Label-wise D & 63.3 & 73.8 & 73.1  & 74.2 & 77.1 & 72.3 \\
		w/o Atlas        & 61.6 & 66.3 & 69.2  & 73.4 & 73.7 & 68.8 \\
		w/o $\mathcal{L}_{\text{SC}}$          & 63.4 & 71.7 & 71.3  & 75.8 & 75.8 & 71.6 \\
		w/o $\mathcal{L}_{\text{CM}}$       & 63.8 & 72.8 & 73.1  & 77.3 & 78.0 & 73.0 \\
		CG-SAMR (proposed)              & \textbf{64.2} & \textbf{75.0} & \textbf{76.4}  & \textbf{80.7} & \textbf{79.1} & \textbf{75.1} \\ \hline
	\end{tabular}
\end{table}

We conduct comprehensive ablation study to separately evaluate the effectiveness of using stretch-out up-sampling module in the decoder network, label-wise discriminators, atlas, lesion shape consistency loss $\mathcal{L}_{\text{SC}}$, and confidence map loss $\mathcal{L}_{\text{CM}}$ in the proposed method. We evaluate each designed module based on two aspects: (1) the effectiveness in data augmentation by the synthesized data, and (2) the contribution on the synthesis quality. For the former, we use the same experimental setting as exp.1 in Table~\ref{tab1}. The effectiveness of modules for data augmentation by synthesis is reported in Table~\ref{tab5}.  Table~\ref{tab6} shows the  contribution of designed modules in the MR image synthesis of different sequences. We can observe that two tables show similar trend. Losing the customized reconstruction for each sequence (stretch-out up-sampling module) can severely degrade the synthesis quality. We find that when atlas is not used in our method, it significantly affects the synthesis quality due to the lack of human brain anatomy prior.  Moreover, dropping either $\mathcal{L}_{\text{SC}}$ or label-wise discriminators in the training also reduces the performance, since the shape consistency loss and the specific supervision on ROIs are not used to optimize the generator to produce more realistic images. In addition, dropping the confidence loss  $\mathcal{L}_{\text{CM}}$ can lead to performance degradation, since the supervision on the intermediate results and attention of uncertain regions during synthesis can provide improved results.

\section{Conclusion}
\label{sec5}
We proposed an effective generative model, called CG-SAMR, for multi-modal MR images, including anatomic $T_1$w, Gd-$T_1$w, $T_2$w, and $FLAIR$, and molecular $APT$w. It was shown that the proposed multi-task optimization under adversarial training further improves the synthesis quality in each ROI. The synthesized data could be used for data augmentation, particularly for images with pathological information of gliomas. Moreover, the proposed approach is an automatic, low-cost solution, which is capable to produce high quality data with diverse content that can be used for training of data-driven methods. We further extended CG-SAMR to UCG-SAMR,  demonstrating the feasibility of using unpaired data for training.

 While our method outperforms state-of-the-art methods to some extent, there are several limitations in our current study. First, all subjects in this study were obtained from a single medical center, so the deep-learning models were not trained and tested on any external data. This leads to the proportional bias in our study without calibration. To make the algorithm more generalizable,
 our future work will incorporate MRI data from multiple external institutions. Second, our method is geared towards synthesizing 2D MR images. Given the lack of the continuity between adjacent scans and cross-sectional analysis, 2D model compromises the fidelity of MR data.   As discussed in section~\ref{sec:da}, along
 the axial direction, the resolution is 4.4 mm for $APT$w images and 1.1 mm for anatomic images. These two non-comparable resolutions limit the application of 3D methods. Moreover, resampling to isotropic for 3D convolution can severely degrade the valuable pathological information in $APT$w images. Therefore, in our future work, the proposed method will be extended to 3D synthesis when comparable quality molecular MRI data is available for training.

\bibliographystyle{IEEEtran}

\begin{thebibliography}{10}
\providecommand{\url}[1]{#1}
\csname url@samestyle\endcsname
\providecommand{\newblock}{\relax}
\providecommand{\bibinfo}[2]{#2}
\providecommand{\BIBentrySTDinterwordspacing}{\spaceskip=0pt\relax}
\providecommand{\BIBentryALTinterwordstretchfactor}{4}
\providecommand{\BIBentryALTinterwordspacing}{\spaceskip=\fontdimen2\font plus
\BIBentryALTinterwordstretchfactor\fontdimen3\font minus
  \fontdimen4\font\relax}
\providecommand{\BIBforeignlanguage}[2]{{%
\expandafter\ifx\csname l@#1\endcsname\relax
\typeout{** WARNING: IEEEtran.bst: No hyphenation pattern has been}%
\typeout{** loaded for the language `#1'. Using the pattern for}%
\typeout{** the default language instead.}%
\else
\language=\csname l@#1\endcsname
\fi
#2}}
\providecommand{\BIBdecl}{\relax}
\BIBdecl

\bibitem{au14}
H.-C. Shin \emph{et~al.}, ``Medical image synthesis for data augmentation and
  anonymization using generative adversarial networks,'' in \emph{International
  workshop on simulation and synthesis in medical imaging}.\hskip 1em plus
  0.5em minus 0.4em\relax Springer, 2018, pp. 1--11.

\bibitem{au8}
P.~Isola, J.-Y. Zhu, T.~Zhou, and A.~A. Efros, ``Image-to-image translation
  with conditional adversarial networks,'' in \emph{Proceedings of the IEEE
  conference on computer vision and pattern recognition}, 2017, pp. 1125--1134.

\bibitem{au9}
T.-C. Wang, M.-Y. Liu, J.-Y. Zhu, A.~Tao, J.~Kautz, and B.~Catanzaro,
  ``High-resolution image synthesis and semantic manipulation with conditional
  gans,'' in \emph{Proceedings of the IEEE Conference on Computer Vision and
  Pattern Recognition}, 2018.

\bibitem{au2}
P.~Y. Wen and S.~Kesari, ``Malignant gliomas in adults,'' \emph{New England
  Journal of Medicine}, vol. 359, no.~5, pp. 492--507, 2008.

\bibitem{au1}
G.~F. Woodworth, T.~Garzon-Muvdi, X.~Ye, J.~O. Blakeley, J.~D. Weingart, and
  P.~C. Burger, ``Histopathological correlates with survival in reoperated
  glioblastomas,'' \emph{Journal of neuro-oncology}, vol. 113, no.~3, pp.
  485--493, 2013.

\bibitem{au62}
P.~Y. Wen \emph{et~al.}, ``Updated response assessment criteria for high-grade
  gliomas: response assessment in neuro-oncology working group,'' \emph{Journal
  of clinical oncology}, vol.~28, no.~11, pp. 1963--1972, 2010.

\bibitem{au60}
J.~Zhou, H.-Y. Heo, L.~Knutsson, P.~C. van Zijl, and S.~Jiang, ``Apt-weighted
  mri: Techniques, current neuro applications, and challenging issues,''
  \emph{Journal of Magnetic Resonance Imaging}, vol.~50, no.~2, pp. 347--364,
  2019.

\bibitem{au3}
S.~Bakas \emph{et~al.}, ``Identifying the best machine learning algorithms for
  brain tumor segmentation, progression assessment, and overall survival
  prediction in the brats challenge,'' \emph{arXiv preprint arXiv:1811.02629},
  2018.

\bibitem{au61}
S.~Tridandapani, ``Radiology “hits refresh” with artificial intelligence,''
  \emph{Academic radiology}, vol.~25, no.~8, pp. 965--966, 2018.

\bibitem{au4}
Y.~Zhou, X.~He, S.~Cui, F.~Zhu, L.~Liu, and L.~Shao, ``High-resolution diabetic
  retinopathy image synthesis manipulated by grading and lesions,'' in
  \emph{International Conference on Medical Image Computing and
  Computer-Assisted Intervention}.\hskip 1em plus 0.5em minus 0.4em\relax
  Springer, 2019, pp. 505--513.

\bibitem{au7}
I.~Goodfellow \emph{et~al.}, ``Generative adversarial nets,'' in \emph{Advances
  in neural information processing systems}, 2014, pp. 2672--2680.

\bibitem{au10}
H.~Van~Nguyen, K.~Zhou, and R.~Vemulapalli, ``Cross-domain synthesis of medical
  images using efficient location-sensitive deep network,'' in
  \emph{International Conference on Medical Image Computing and
  Computer-Assisted Intervention}.\hskip 1em plus 0.5em minus 0.4em\relax
  Springer, 2015, pp. 677--684.

\bibitem{au12}
A.~Chartsias, T.~Joyce, M.~V. Giuffrida, and S.~A. Tsaftaris, ``Multimodal mr
  synthesis via modality-invariant latent representation,'' \emph{IEEE
  transactions on medical imaging}, vol.~37, no.~3, pp. 803--814, 2017.

\bibitem{au13}
N.~Cordier, H.~Delingette, M.~L{\^e}, and N.~Ayache, ``Extended modality
  propagation: image synthesis of pathological cases,'' \emph{IEEE transactions
  on medical imaging}, vol.~35, no.~12, pp. 2598--2608, 2016.

\bibitem{au59}
P.~Guo, P.~Wang, J.~Zhou, V.~M. Patel, and S.~Jiang, ``Lesion mask-based
  simultaneous synthesis of anatomic and molecularmr images using a gan,''
  \emph{arXiv preprint arXiv:2006.14761}, 2020.

\bibitem{au23}
A.~F. Frangi, S.~A. Tsaftaris, and J.~L. Prince, ``Simulation and synthesis in
  medical imaging,'' \emph{IEEE transactions on medical imaging}, vol.~37,
  no.~3, pp. 673--679, 2018.

\bibitem{au24}
Y.~Huo \emph{et~al.}, ``Synseg-net: Synthetic segmentation without target
  modality ground truth,'' \emph{IEEE transactions on medical imaging},
  vol.~38, no.~4, pp. 1016--1025, 2018.

\bibitem{au25}
S.~Roy, A.~Carass, and J.~L. Prince, ``Magnetic resonance image example-based
  contrast synthesis,'' \emph{IEEE transactions on medical imaging}, vol.~32,
  no.~12, pp. 2348--2363, 2013.

\bibitem{au27}
S.~Roy, A.~Carass, and J.~Prince, ``A compressed sensing approach for mr tissue
  contrast synthesis,'' in \emph{Biennial International Conference on
  Information Processing in Medical Imaging}.\hskip 1em plus 0.5em minus
  0.4em\relax Springer, 2011, pp. 371--383.

\bibitem{au26}
A.~Jog, A.~Carass, S.~Roy, D.~L. Pham, and J.~L. Prince, ``Random forest
  regression for magnetic resonance image synthesis,'' \emph{Medical image
  analysis}, vol.~35, pp. 475--488, 2017.

\bibitem{au28}
Y.~Huang, L.~Beltrachini, L.~Shao, and A.~F. Frangi, ``Geometry regularized
  joint dictionary learning for cross-modality image synthesis in magnetic
  resonance imaging,'' in \emph{International Workshop on Simulation and
  Synthesis in Medical Imaging}.\hskip 1em plus 0.5em minus 0.4em\relax
  Springer, 2016, pp. 118--126.

\bibitem{au29}
M.~I. Miller, G.~E. Christensen, Y.~Amit, and U.~Grenander, ``Mathematical
  textbook of deformable neuroanatomies,'' \emph{Proceedings of the National
  Academy of Sciences}, vol.~90, no.~24, pp. 11\,944--11\,948, 1993.

\bibitem{au30}
M.~J. Cardoso, C.~H. Sudre, M.~Modat, and S.~Ourselin, ``Template-based
  multimodal joint generative model of brain data,'' in \emph{International
  conference on information processing in medical imaging}.\hskip 1em plus
  0.5em minus 0.4em\relax Springer, 2015, pp. 17--29.

\bibitem{au35}
R.~Li \emph{et~al.}, ``Deep learning based imaging data completion for improved
  brain disease diagnosis,'' in \emph{International Conference on Medical Image
  Computing and Computer-Assisted Intervention}.\hskip 1em plus 0.5em minus
  0.4em\relax Springer, 2014, pp. 305--312.

\bibitem{au36}
H.~Van~Nguyen, K.~Zhou, and R.~Vemulapalli, ``Cross-domain synthesis of medical
  images using efficient location-sensitive deep network,'' in
  \emph{International Conference on Medical Image Computing and
  Computer-Assisted Intervention}.\hskip 1em plus 0.5em minus 0.4em\relax
  Springer, 2015, pp. 677--684.

\bibitem{au34}
V.~Sevetlidis, M.~V. Giuffrida, and S.~A. Tsaftaris, ``Whole image synthesis
  using a deep encoder-decoder network,'' in \emph{International Workshop on
  Simulation and Synthesis in Medical Imaging}.\hskip 1em plus 0.5em minus
  0.4em\relax Springer, 2016, pp. 127--137.

\bibitem{au37}
D.~Nie \emph{et~al.}, ``Medical image synthesis with context-aware generative
  adversarial networks,'' in \emph{International Conference on Medical Image
  Computing and Computer-Assisted Intervention}.\hskip 1em plus 0.5em minus
  0.4em\relax Springer, 2017, pp. 417--425.

\bibitem{au38}
H.~R. Roth \emph{et~al.}, ``Spatial aggregation of holistically-nested
  convolutional neural networks for automated pancreas localization and
  segmentation,'' \emph{Medical image analysis}, vol.~45, pp. 94--107, 2018.

\bibitem{au39}
J.-Y. Zhu, T.~Park, P.~Isola, and A.~A. Efros, ``Unpaired image-to-image
  translation using cycle-consistent adversarial networks,'' in \emph{Computer
  Vision (ICCV), 2017 IEEE International Conference on}, 2017.

\bibitem{au48}
M.-Y. Liu, T.~Breuel, and J.~Kautz, ``Unsupervised image-to-image translation
  networks,'' in \emph{Advances in neural information processing systems},
  2017, pp. 700--708.

\bibitem{au40}
J.~M. Wolterink, A.~M. Dinkla, M.~H. Savenije, P.~R. Seevinck, C.~A. van~den
  Berg, and I.~I{\v{s}}gum, ``Mr-to-ct synthesis using cycle-consistent
  generative adversarial networks,'' in \emph{Proc. Neural Inf. Process.
  Syst.(NIPS)}, 2017.

\bibitem{au41}
A.~Chartsias, T.~Joyce, R.~Dharmakumar, and S.~A. Tsaftaris, ``Adversarial
  image synthesis for unpaired multi-modal cardiac data,'' in
  \emph{International workshop on simulation and synthesis in medical
  imaging}.\hskip 1em plus 0.5em minus 0.4em\relax Springer, 2017, pp. 3--13.

\bibitem{au42}
Z.~Zhang, L.~Yang, and Y.~Zheng, ``Translating and segmenting multimodal
  medical volumes with cycle-and shape-consistency generative adversarial
  network,'' in \emph{Proceedings of the IEEE conference on computer vision and
  pattern Recognition}, 2018, pp. 9242--9251.

\bibitem{au43}
Y.~Huo, Z.~Xu, S.~Bao, A.~Assad, R.~G. Abramson, and B.~A. Landman,
  ``Adversarial synthesis learning enables segmentation without target modality
  ground truth,'' in \emph{2018 IEEE 15th International Symposium on Biomedical
  Imaging (ISBI 2018)}.\hskip 1em plus 0.5em minus 0.4em\relax IEEE, 2018, pp.
  1217--1220.

\bibitem{au44}
A.~Kendall and Y.~Gal, ``What uncertainties do we need in bayesian deep
  learning for computer vision?'' in \emph{Advances in neural information
  processing systems}, 2017, pp. 5574--5584.

\bibitem{au45}
A.~Kendall, Y.~Gal, and R.~Cipolla, ``Multi-task learning using uncertainty to
  weigh losses for scene geometry and semantics,'' in \emph{Proceedings of the
  IEEE conference on computer vision and pattern recognition}, 2018, pp.
  7482--7491.

\bibitem{au46}
R.~Yasarla and V.~M. Patel, ``Confidence measure guided single image
  de-raining,'' \emph{IEEE Transactions on Image Processing}, vol.~29, pp.
  4544--4555, 2020.

\bibitem{au51}
J.~M.~J. Valanarasu, R.~Yasarla, P.~Wang, I.~Hacihaliloglu, and V.~M. Patel,
  ``Learning to segment brain anatomy from 2d ultrasound with less data,''
  \emph{IEEE Journal of Selected Topics in Signal Processing}, 2020.

\bibitem{au15}
J.~Johnson, A.~Alahi, and L.~Fei-Fei, ``Perceptual losses for real-time style
  transfer and super-resolution,'' in \emph{European conference on computer
  vision}.\hskip 1em plus 0.5em minus 0.4em\relax Springer, 2016, pp. 694--711.

\bibitem{au49}
K.~He, X.~Zhang, S.~Ren, and J.~Sun, ``Deep residual learning for image
  recognition,'' in \emph{Proceedings of the IEEE conference on computer vision
  and pattern recognition}, 2016, pp. 770--778.

\bibitem{au50}
R.~Yasarla and V.~M. Patel, ``Uncertainty guided multi-scale residual
  learning-using a cycle spinning cnn for single image de-raining,'' in
  \emph{Proceedings of the IEEE Conference on Computer Vision and Pattern
  Recognition}, 2019, pp. 8405--8414.

\bibitem{au11}
O.~Ronneberger, P.~Fischer, and T.~Brox, ``U-net: Convolutional networks for
  biomedical image segmentation,'' in \emph{International Conference on Medical
  image computing and computer-assisted intervention}.\hskip 1em plus 0.5em
  minus 0.4em\relax Springer, 2015, pp. 234--241.

\bibitem{au16}
C.~H. Sudre, W.~Li, T.~Vercauteren, S.~Ourselin, and M.~J. Cardoso,
  ``Generalised dice overlap as a deep learning loss function for highly
  unbalanced segmentations,'' in \emph{Deep learning in medical image analysis
  and multimodal learning for clinical decision support}.\hskip 1em plus 0.5em
  minus 0.4em\relax Springer, 2017, pp. 240--248.

\bibitem{au5}
S.~Jiang \emph{et~al.}, ``Identifying recurrent malignant glioma after
  treatment using amide proton transfer-weighted mr imaging: a validation study
  with image-guided stereotactic biopsy,'' \emph{Clinical Cancer Research},
  vol.~25, no.~2, pp. 552--561, 2019.

\bibitem{au17}
Y.~Zhang \emph{et~al.}, ``Selecting the reference image for registration of
  cest series,'' \emph{Journal of Magnetic Resonance Imaging}, vol.~43, no.~3,
  pp. 756--761, 2016.

\bibitem{au20}
J.~Lipkova \emph{et~al.}, ``Personalized radiotherapy design for glioblastoma:
  integrating mathematical tumor models, multimodal scans, and bayesian
  inference,'' \emph{IEEE transactions on medical imaging}, vol.~38, no.~8, pp.
  1875--1884, 2019.

\bibitem{au18}
N.~J. Tustison \emph{et~al.}, ``N4itk: improved n3 bias correction,''
  \emph{IEEE transactions on medical imaging}, vol.~29, no.~6, pp. 1310--1320,
  2010.

\bibitem{au19}
L.~G. Ny{\'u}l, J.~K. Udupa, and X.~Zhang, ``New variants of a method of mri
  scale standardization,'' \emph{IEEE transactions on medical imaging},
  vol.~19, no.~2, pp. 143--150, 2000.

\bibitem{au21}
B.~B. Avants, N.~J. Tustison, G.~Song, P.~A. Cook, A.~Klein, and J.~C. Gee, ``A
  reproducible evaluation of ants similarity metric performance in brain image
  registration,'' \emph{Neuroimage}, vol.~54, no.~3, pp. 2033--2044, 2011.

\bibitem{au52}
J.~Ashburner \emph{et~al.}, ``Spm12 manual,'' \emph{Wellcome Trust Centre for
  Neuroimaging, London, UK}, p. 2464, 2014.

\end{thebibliography}
\end{document}